\documentclass[]{elsart}

\usepackage{epsfig}

\usepackage{amssymb}

\begin{document}

\begin{frontmatter}



\title{Monte-Carlo simulations of the background of the coded-mask camera for
  X- and Gamma-rays on-board the Chinese-French GRB mission SVOM}

\author[leicester]{O. Godet\corauthref{cor}\thanksref{label2}},
\corauth[cor]{Corresponding author.}
\ead{og19@star.le.ac.uk}
\thanks[label2]{Present address: CESR - 9 avenue du Colonel Roche, 31028
  Toulouse Cedex 4, France. Tel: +33 561558371; Fax: +33 561556701}
\author[irfu]{P. Sizun}
\author[cesr]{, D. Barret}
\author[cesr]{, P. Mandrou}
\author[irfu2]{ B. Cordier}
\author[irfu2]{, S. Schanne}
\author[cesr]{\& N. Remou\'e}

\address[leicester]{X-ray and Observational Astronomy Group, Department of
Physics \& Astronomy, University of Leicester, LE1 7RH, UK}

\address[irfu]{CEA, IRFU, SEDI, 91191 Gif-sur-Yvette, France}

\address[cesr]{Centre d'Etude Spatiale des Rayonnements, 9 avenue du Colonel
Roche, 31047 Toulouse, France}

\address[irfu2]{CEA, IRFU, Service d'Astrophysique, 91191 Gif-sur-Yvette, France}

\author{}

\address{}

\begin{abstract}
% Text of abstract

For several decades now, wide-field coded mask cameras have been used with
  success to localise Gamma-ray bursts (GRBs). In these instruments, the event
  count rate is dominated by the photon background due to their large field of
  view and large effective area. It is therefore essential to estimate the
  instrument background expected in orbit during the early phases of the
  instrument design in order to optimise the scientific performances of the
  mission.  We present here a detailed study of the instrument background and
  sensitivity of the coded-mask camera for X- and Gamma-rays (CXG) to be used
  in the detection and localisation of high-redshift GRBs on-board the
  international GRB mission SVOM.  To compute the background spectrum, a
  Monte-Carlo approach was used to simulate the primary and secondary
  interactions between particles from the main components of the space
  environment that SVOM will encounter along its Low Earth Orbit (LEO) (with
  an altitude of 600 km and an inclination of $\sim 30^\circ$) and the body of
  the CXG. We consider the detailed mass model of the CXG in its latest
  design.  According to our results, i) the design of the passive shield of
  the camera ensures that in the 4-50 keV imaging band the cosmic X-/Gamma-ray
  background is dominant whilst the internal background should start to become
  dominant above 70-90 keV; ii) the current camera design ensures that the CXG
  camera will be more sensitive to high-redshift GRBs than the {\it Swift}
  Burst Alert Telescope thanks to a low-energy threshold of 4\,keV.

\end{abstract}

\begin{keyword}
Monte-carlo simulation \sep background \sep Hard X-rays 
\PACS 
\end{keyword}
\end{frontmatter}

% main text

\section{Introduction}
\label{payload}
\label{imager}

Gamma-ray bursts (GRBs) are highly transient and powerful cosmological events
appearing in the sky for very short times (from a few milli-seconds to
hundreds of seconds). They are considered to be associated to the death of
massive stars \cite{MacFayden} or compact object mergers \cite{Eichler}.
Past and current space missions dedicated to the study of GRBs have
demonstrated that the most efficient way to detect and localise them in the
hard X-ray and Gamma-ray domain is to use wide-field coded-mask cameras.

\begin{figure*}
\begin{center}
\begin{tabular}{c}
\hspace{-1.5cm}\psfig{figure=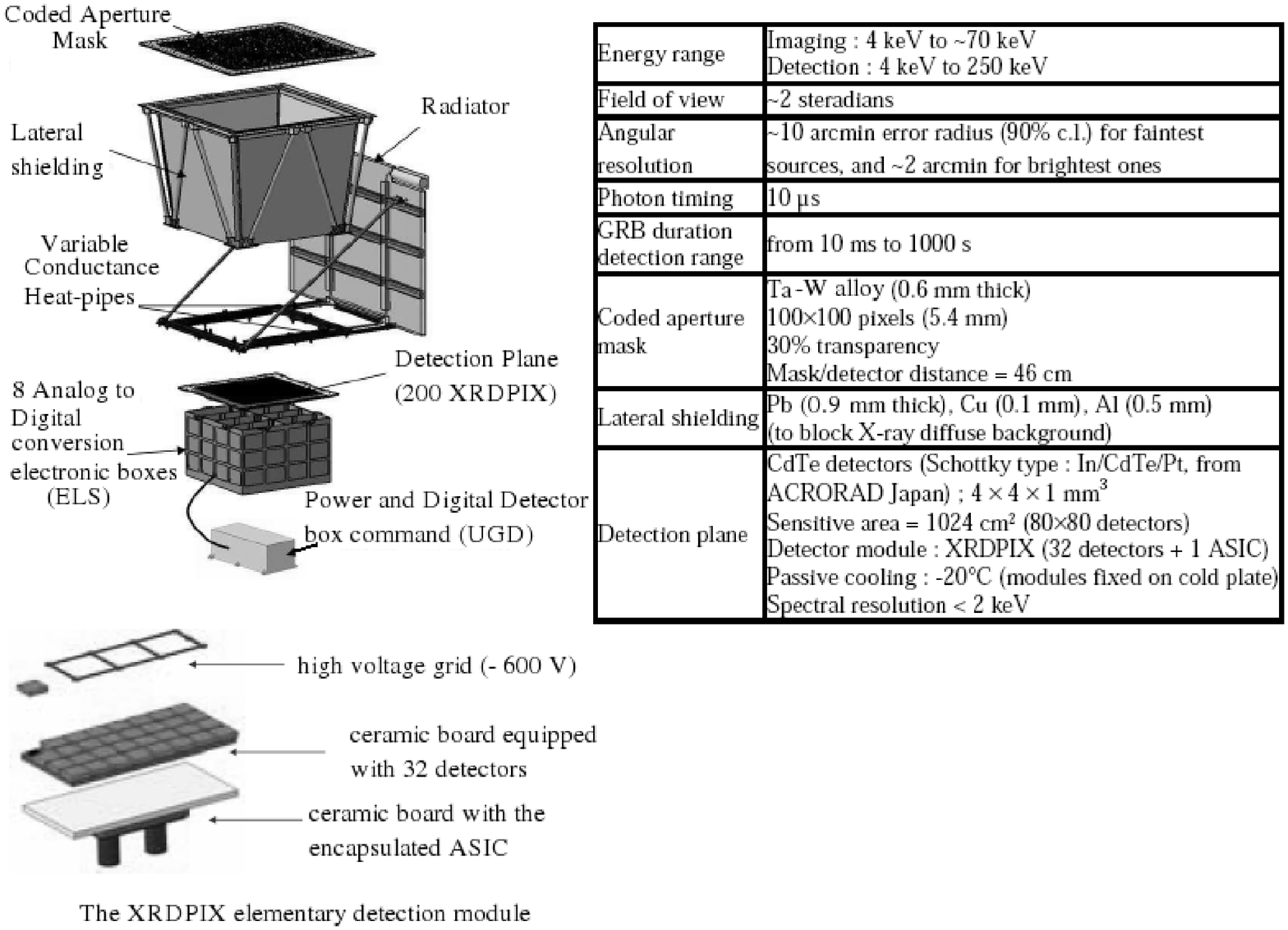,width=17cm} \\
\end{tabular}
\caption{View of ECLAIRs main components and summary of ECLAIRs main characteristics.}
\label{figure1}
\end{center}
\end{figure*}

We present here the main instrument of the science payload on-board the
international GRB mission SVOM (Space Variable Object Monitor)
\cite{Schanne2,Schanne3}, the coded-mask camera for X- and Gamma-rays (CXG)
responsible for triggering GRB observations in the 4-250 keV energy band and
their localisation with an accuracy better than 10 arc-minutes in the 4-50 keV
imaging band. SVOM, which is the evolution of the mission {\it ECLAIRs}
\cite{Barret, Schanne} from micro- to mini-satellite, is dedicated to the
study of high-redshift GRBs. It is expected to be launched in 2013. The CXG
has a large field of view (FoV $\sim$ 2 sr). The CXG passive shielding is
designed so that the background photons coming from outside the FoV will be
completely absorbed by the compounds of the shielding (see Section~\ref{SVOM}
and Fig.~\ref{figure1}).  The detection plane, DPIX \cite{Ehanno} with an
effective area of 1024 cm$^2$ is an assembly of 200 elementary modules
(XRDPIX) equipped with 32 CdTe Schottky detectors ($4\times 4$\,mm$^2$, 1\,mm
thickness) produced by ACRORAD Co. LTD in Japan (e.g. \cite{Takahashi}).  Each
XRDPIX is read out by the very low noise front-end ASIC IDeF-X \cite{Gevin},
which will enable the camera to reach a low-energy threshold of $\sim 4$\,keV
after careful selection of the 6400 CdTe detectors of the detection plane (see
Fig.~\ref{figure1}). First lab measurements using a CdTe detector coupled with
IDeF-X reported a low-energy threshold around 2.7 keV (see Fig.\,4 in
\cite{Ehanno}; see also \cite{Remoue}). This is a significant technological
improvement when compared to the 12-15 keV low-energy threshold of the {\it
Swift} BAT (Burst Alert Telescope) or INTEGRAL/ISGRI (Integral Soft Gamma-Ray
Imager) using similar detectors. The CXG is currently in a B phase at the CNES
(Centre National d'Etudes Spatiales, the French Space Agency).

%===============================================

Coded-mask cameras such as the CXG are known to be photon background-dominated
instruments due to their large FoV and effective area. Even if GRBs have high
signal-to-noise ratios, it is still of primary importance to evaluate the
camera background expected in orbit during the early phases of the design in
order to optimise the instrument capability to observe GRBs as well as non GRB
targets.  This paper focuses on the estimation of the background spectrum of
the CXG camera, depending on its design and mission parameters, and the impact
on the scientific performances of the mission.  This paper is organised as
follows:
\begin{itemize} 
\item[{\bf \S 2}] We give a brief description of the Monte-Carlo simulator,
Geant, as well as the main components of the space environment that the
spacecraft will encounter once in orbit, used to perform our simulations. We
also define the mass model of the CXG camera used to perform our simulations.

\item[{\bf \S 3}] We discuss the main features of the background spectrum.  We
also compute the camera limiting sensitivity for GRBs, and we compare it with
those obtained with other GRB trigger instruments. We investigate what
impact on the camera performance would have the existence of a dead layer on
the detectors.

\item[{\bf \S 4}] We present the main conclusions of the paper and discuss
further improvements.
\end{itemize}

\section{Monte-Carlo simulations of the instrument background}

\subsection{Simulation tool}

The simulations were performed using the CERN Monte-Carlo code
Geant\,\footnote{http://geant4.web.cern.ch/geant4}, initially designed to
model the interactions between matter and particles in high-energy nuclear
physics experiments. Geant has since been used in the astrophysical field to
model the performance of space instruments such as SPI (SPectrometer on
INTEGRAL) on-board INTEGRAL \cite{Sturner} and Fermi/GLAST \cite{Kippen}, for
instance.

Geant enables us to: i) describe the detailed mass model of the payload and
the spacecraft; ii) draw particles (photons, electrons, protons, neutrons,
...)  following a specific spatial and energy distribution; iii) track the
paths of the primary particles through the body of the camera and the
spacecraft as well as any secondary particles generated during the different
physical processes such as Compton, Rayleigh scattering, photo-electric
effects, pair annihilation and creation, nuclear interaction.

The simulations were performed using the release 4.9.1 of the Geant4 C$++$
 toolkit \cite{Agostinelli} along with the low energy electromagnetic physics
 dataset.

\subsection{Space environment and GRB model}
\label{environment}

SVOM is a LEO mission with an altitude of 600 km and an inclination of
$30^\circ$ \cite{Schanne2,Schanne3}. The satellite will then be subject to
different sources of background in space: extragalactic components (X-ray and
Gamma-ray diffuse background, primary proton and electron cosmic rays) and
near Earth components (Gamma-ray albedo, neutrons, secondary protons located
under the radiation belt).  At the moment, we only take into account the
extragalactic diffuse background, because it is the most relevant source of
background for the study presented here (see Section~\ref{bkgs}).

The main source of background radiation for a wide-field camera is the
quasi-isotropic cosmic X-/Gamma-ray background \cite{Giacconi}.  The spectrum
of the cosmic background that we used is from \cite{gruber} (from 10 keV to 2
MeV). \cite{Ajello} showed that the {\it Swift}-BAT agrees within 8\% with the
normalisation of the cosmic background spectrum given in \cite{gruber} below 2
MeV.

The cosmic background spectrum measured by {\it XMM-Newton}
\cite{DeLuca}, RXTE \cite{Revnivtsev} and {\it Chandra} \cite{Hickox} seems to
be larger by 25-40\% in the 2-10 keV energy range when compared to the
spectrum given in \cite{gruber}.  \cite{Frontera} argued that the
discrepancies between the different instruments in this energy band may be
related to systematic errors in the response function used for diffuse
sources.  However, this discrepancy needs to be confirmed (see
\cite{Moretti}). 

For our simulations, we decided to extend the shape of the cosmic background
spectrum given in \cite{gruber} below 10 keV.  We considered a $2\,\pi$ sr
spatial distribution for the cosmic background photons in our simulations
since the Earth acts like a screen on the other hemisphere.

\subsubsection{GRB model}

GRBs are assumed to be point-like sources with a spectral distribution given
by the Band's function \cite{Band} as follows:
\[N(\rm E) = A_0 \left\{
\begin{array}{ll}
\left(\frac{E}{100\,\rm keV}\right)^\alpha\,\exp\left(-\frac{\rm E}{\rm
E_0 }\right), & \rm E \leq \rm E_p \\ 

\left(\frac{E_0}{100\,\rm
keV}\right)^{(\alpha-\beta)}\,{\rm e}^{(\beta-\alpha)}\,\left(\frac{\rm E}{\rm
100\,\rm keV}\right)^{\beta}, & \rm E > \rm E_p \\
\end{array}
\right. \] where $A_0$ is in units of photons cm$^{-2}$ s$^{-1}$ keV$^{-1}$
 and the peak energy $E_p = (\alpha - \beta)~E_0$ is the energy for which the
 radiated energy reaches a maximum.

\subsection{Mass model of the CXG camera}
\label{SVOM}

Below we describe the specifications of the mass model of the CXG camera used
to perform the simulations following the mass and dimension restrictions of
the SVOM mission. Table~\ref{tab1} summarises the main characteristics of the
camera.

\begin{table*}[h]
\caption{Summary of the main characteristics of the CXG camera as designed for
the mission SVOM as well as the mission details.}
\label{tab1}
\begin{center}
\begin{tabular}{|l|c|}
\hline
{\bf Orbit:}  &   Altitude = 600 km \\
        &   Inclination = $30^\circ$ \\
\hline
{\bf Camera:} & \\

Mass of the camera &   $\sim 63$ kg  \\
Mass of the platform &    $300$ kg \\
Mask-detector plane height &  46 cm\\
\hline

{\bf Detection plane:}                   &    \\
Number of pixels  &  $80\times 80$\\
Detector surface  &  $4\times 4$ mm$^{2}$ \\
Detector thickness &  1 mm \\
Efficient surface &  $1024$ cm$^{2}$ \\
\hline
{\bf Passive shielding:}  & \\
Compounds$^{+}$  & Pb (0.9 mm), Cu (0.1 mm), \\
           & Al (0.5 mm) \\

Absorption in 4-50 keV$^*$ & 100\% \\
% \hline
%\end{tabular}
%\end{center}
%\end{table*}

%\begin{table*}[h]
%\caption{The second part of Table~\ref{tab1}.}
%\begin{center}
%\begin{tabular}{|l|c|}

\hline
{\bf Coded mask:}  &  \\
Aperture fraction  &  30\% \\
Compounds  &    TaW alloy \\
           & (97.5\% Ta \& 2.5\%W)$^\dagger$\\
           &  MLI layer$^\ddagger$ \\                                  
           &  50 $\mu$m of Kapton, 12 $\mu$m of Mylar  \\
           &  0.5 $\mu$m of Al \\
%Absorption in 4-50 keV$^*$ &  $> 95\%$   &   \\
\hline
{\bf Field of view:}  & \\
Total FoV    & $88.7^\circ\times 88.7^\circ$ (2.04 sr) \\
totally coded FoV    &  $22.1^\circ\times 22.1^\circ$ (0.15 sr) \\

\hline
\end{tabular}
    \begin{list}{}{}
      \item $^+$ The compounds are given from the outside to the inside. 
      \item $^*$ The imaging energy band of the CXG camera.
      \item $^\dagger$ The percentages are given as a fraction of the 2.9 kg
      coded-mask mass.
      \item $^\ddagger$ The MLI layer will be put on top of the coded mask to
      prevent optical photon loading on the detection plane.
    \end{list}
\end{center}
\end{table*}

\begin{figure*}
\begin{center}
\begin{tabular}{cc}
\hspace{-1.5cm}\psfig{figure=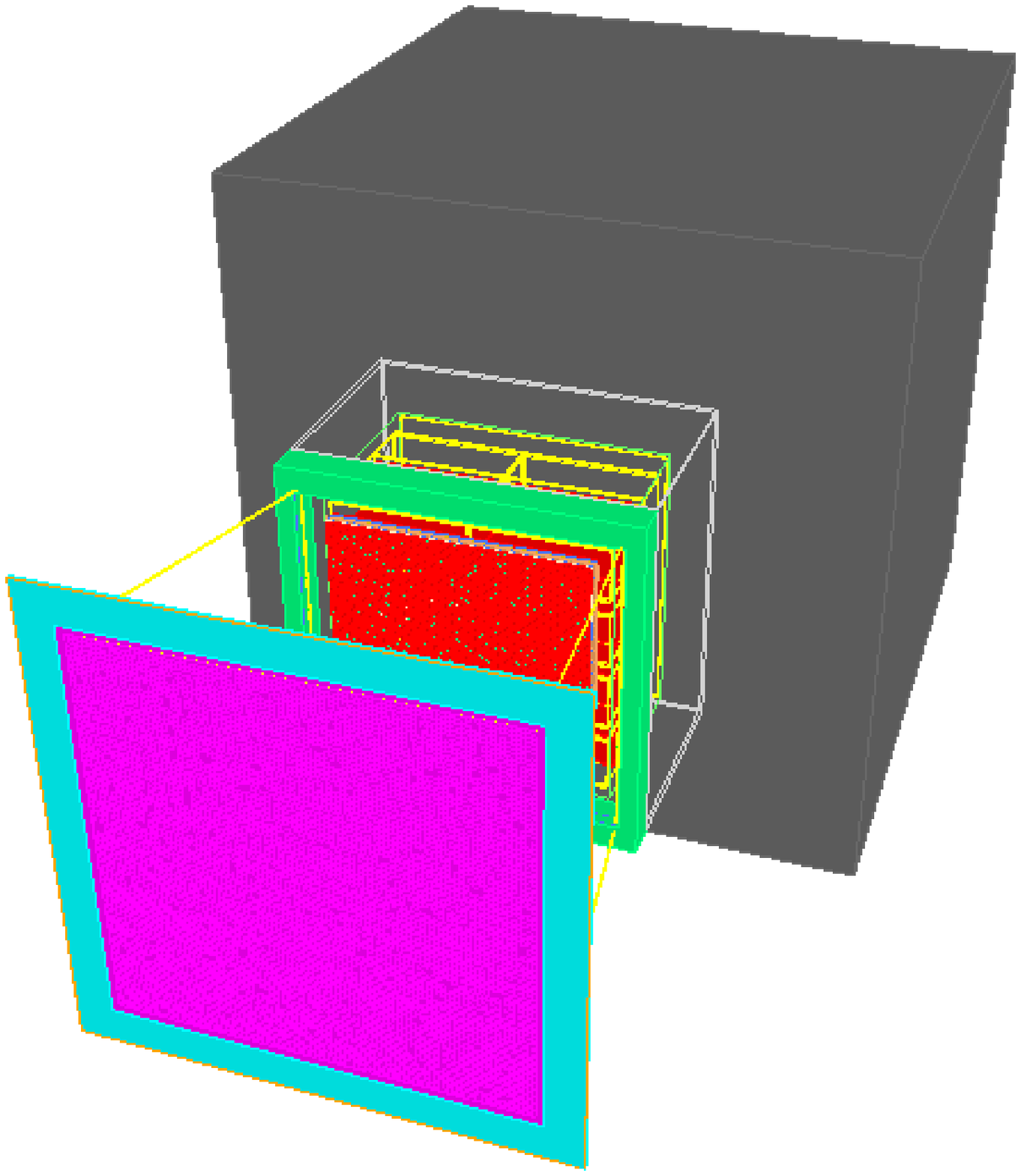,width=7cm} & \psfig{figure=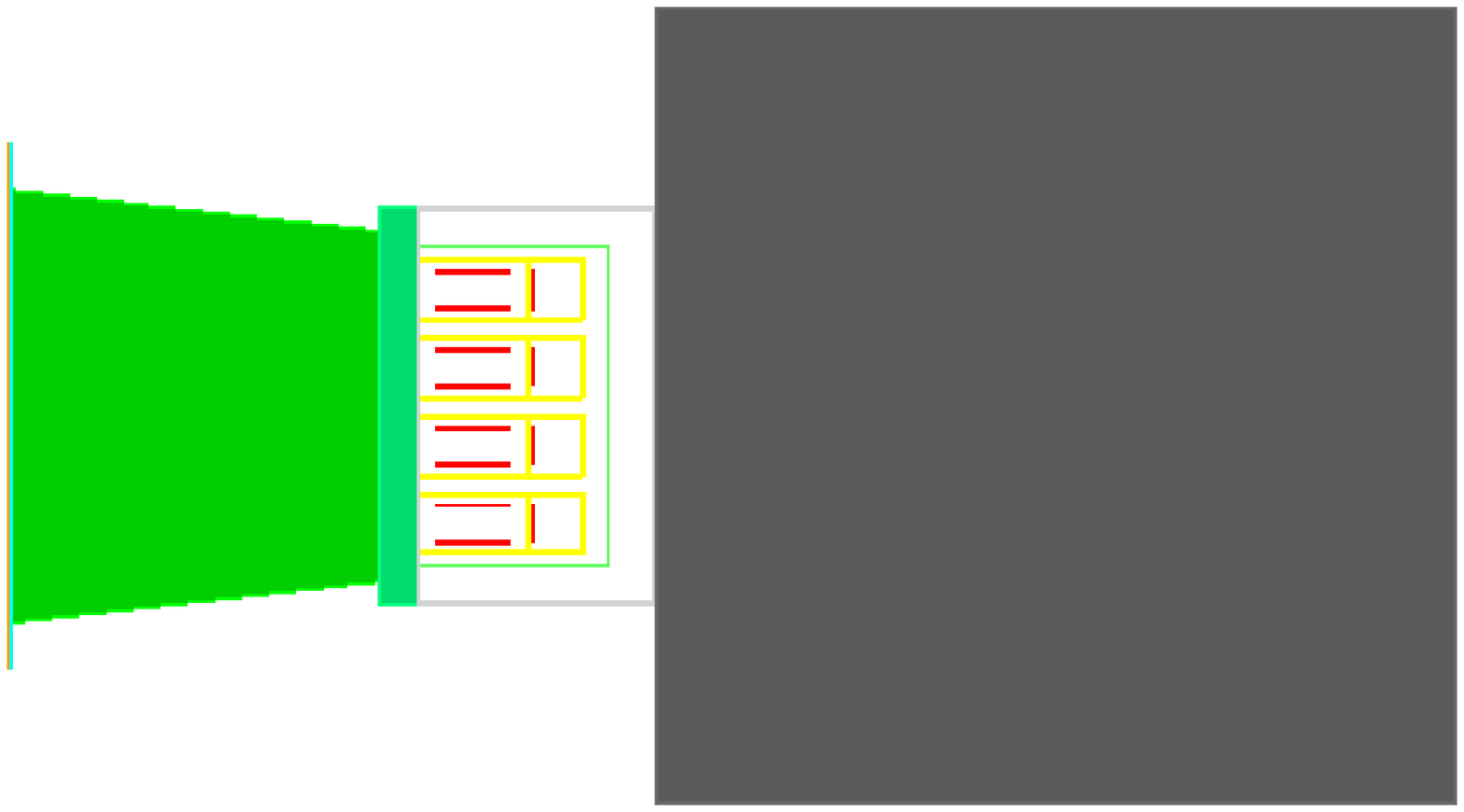,width=9cm} \\
\end{tabular}
\caption{Scheme of the mass model of the CXG as well as the cold plate and the
readout electronics box placed under the camera, as designed for the mission
SVOM (see Section~\ref{SVOM}). The grey box corresponds to the satellite
platform.}
\label{fig_ECLAIRs}
\end{center}
\end{figure*}

On-board the SVOM mission (see Fig.~\ref{fig_ECLAIRs}), the detection plane
consists of $80~\times{}~80$ 1\,mm-thick CdTe pixels with a useful width of
4~mm, leading to an effective area of $1024~\mathrm{cm}^2$. The ceramics
supporting the detector and the ASIC are taken into account in the
simulations. However, the high voltage grid is not modelled. Made of a Ta-W
alloy, the $500~\mu{}m$ thick and 54~cm wide coded mask, whose presently
square pattern contains 30 percent open cells, is located about 46~cm above
the detectors, which provides the instrument with a field of view of 2~sr.
The mass model also includes a multi-layer thermal coating insulation (MLI)
above the coded mask to prevent optical photon loading on the detection plane,
the TA6V upper- and lower supports of the mask, as well as simplified versions
of the cold plate in AlBeMet and the readout electronics box.

After simulations, we chose a graded shield combining Pb, Cu and Al, which is a
trade-off between a maximum reduction of the background in the 4-50~keV energy
band, mass budget considerations and the necessity to keep a few instrumental
gamma-ray lines at higher energies for calibration purposes, especially lead
K$\alpha$ and K$\beta$ fluorescence lines.

The remainder of the satellite is assumed to be a $1~\mathrm{m}^3$ 300~kg
cube of pseudo-Aluminium, which has an averaged density defined as the ratio
of the spacecraft mass over its volume, the exact structure of the spacecraft
being not yet known with accuracy.

\section{Results on the background level and the instrument sensitivity}
\label{result}

\subsection{The background spectrum}
\label{bkgs}

\begin{figure}
\begin{center}
\hspace{1cm}\psfig{figure=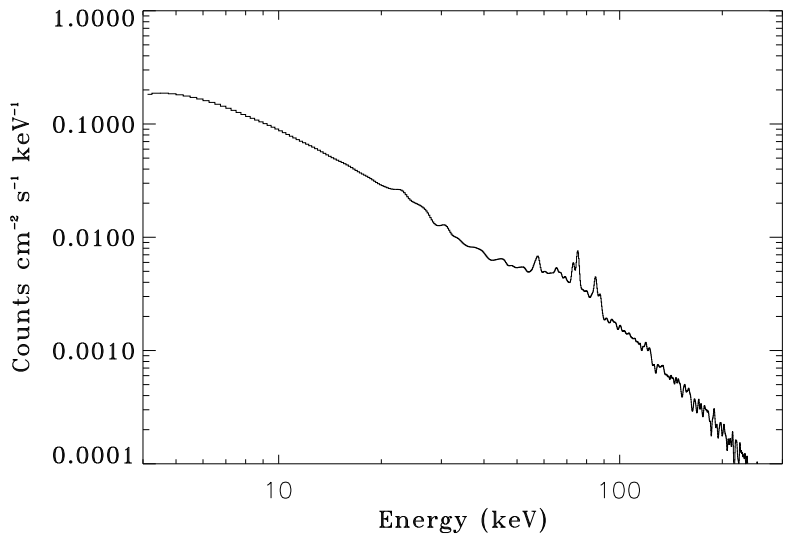,width=11cm} 
\caption{Averaged spectrum of the background induced by the X-ray diffuse
  background (outside the South Atlantic Anomaly) on the detection plane of
  the CXG. The spectrum is degraded to the resolution of the CdTe detectors
  using a Gaussian with a FWHM of 1.6 keV as measured in the CESR lab
  facility. }
\label{fig_bkg_ECLAIRs}
\end{center}
\end{figure}

Fig.~\ref{fig_bkg_ECLAIRs} shows the spectrum of the background measured on
the detection plane resulting from our Monte-Carlo simulations in the 4
keV-300 keV energy range. The spectrum is degraded to the resolution of the
CdTe detectors using a Gaussian with a FWHM of $\sim 1.6$ keV as measured in
the CESR lab facility.

\begin{figure*}
\begin{center}
\hspace{1.cm}\psfig{figure=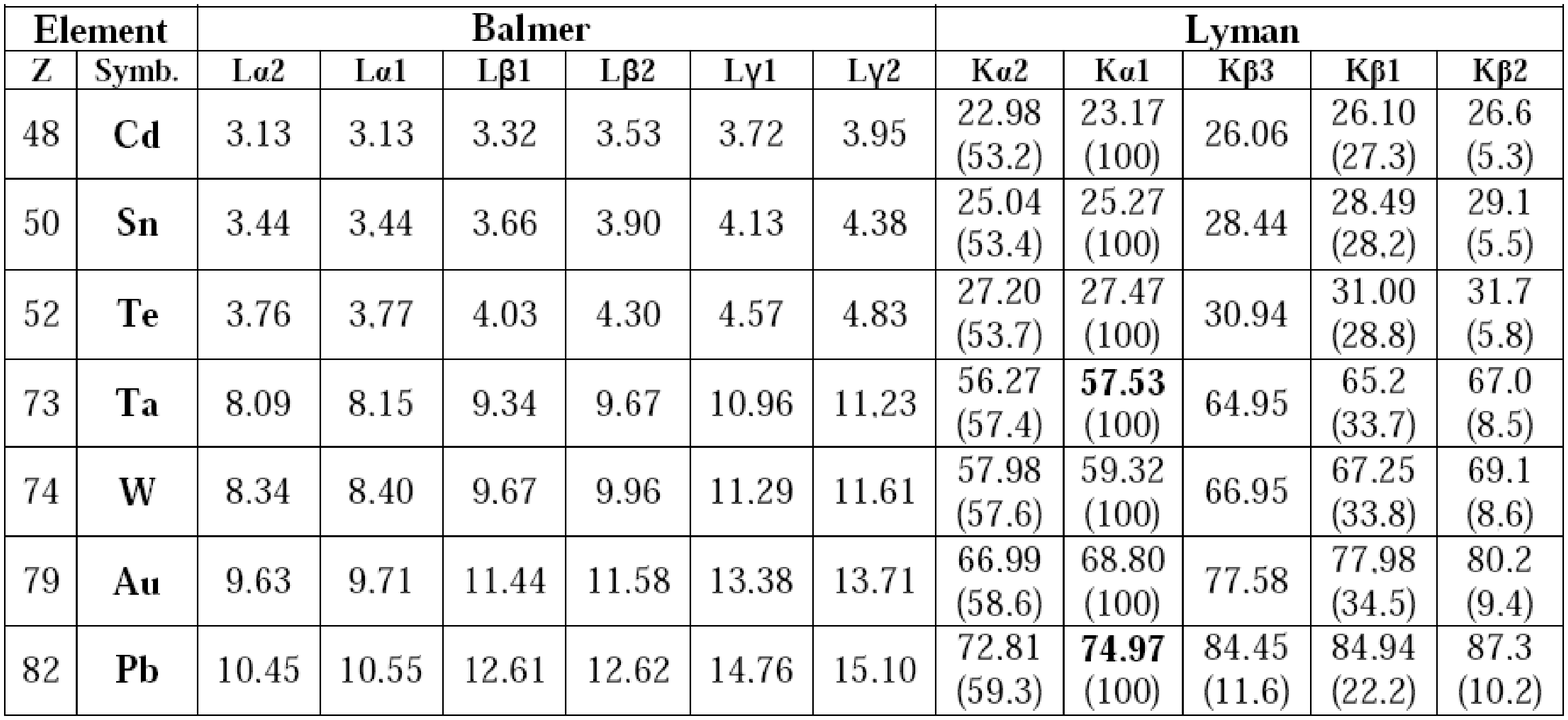,width=13cm} 
\caption{Summary of the fluorescence lines observed in the background
  spectra. The numbers mentioned in parenthesis correspond to the relative
  intensities of the lines \cite{Lederer}. The line energies come from
  \cite{Deslattes}. Note that the values mentioned in the Table are slightly
  different from those used in Geant4 \cite{Guatelli}.}
\label{fig_tab2}
\end{center}
\end{figure*}

The presence of the MLI layer on top of the coded mask results in a decrease
of the background level at low energy, since it absorbs a fraction of the
incident low-energy photons. Thus, the MLI layer is responsible for decreasing
the background level by $\sim 11\%$, in the 4-50 keV energy band (see
Fig.~\ref{fig_bkg_deadlayer}).  The compounds and thicknesses of the passive
shielding and the coded mask, as summarised in Table~\ref{tab1}, lead to the
apparition of several fluorescence lines (see Fig.~\ref{fig_bkg_ECLAIRs} and
Fig.~\ref{fig_tab2}; the line characteristics in Fig.~\ref{fig_tab2} are from
\cite{Lederer,Deslattes,Guatelli}). The intensity of the Au fluorescence lines
coming from the electronics is likely to be overestimated due to the crude
modelling of the electronics in our simulations at the moment. K$\alpha$ and
K$\beta$ fluorescence lines from the Cd and Te are also seen in the background
spectrum. The most intense instrumental lines (Pb-Ta lines) will be used to
calibrate the gain of the detection plane in orbit. To ensure a better control
on the gain over the energy band of the CXG camera (4-250 keV), we are
investigating the benefit of considering some fluorescence lines at low energy
(for instance, Cu K$\alpha$ 8.04 keV) without degrading too much the
scientific performances of the camera.  From the simulations, we checked that
the efficiency of the passive shielding to block the photon background outside
the camera field of view is 100\% in the 4-50 keV imaging band as
required. The averaged background count rate in the 4-50 keV imaging band is
1.7 counts cm$^{-2}$ s$^{-1}$.

In addition to the external background (the X-ray diffuse background), the
camera will present an internal background mainly due to charged particles
interacting with the body of the camera. Even if we did not yet perform
detailled simulations to model this background component, we can estimate that
the CXG background should be dominated in the 4-50 keV imaging band (i.e. the
relevant energy band here) by the X-ray diffuse background outside the SAA
(South Atlantic Anomaly).  Simulations outside the SAA on a previous design of
the CXG camera to be embarked on-board the micro-satellite {\it ECLAIRs}
showed that the transition between the internal (due to primary and secondary
protons, primary electrons, neutrons and gamma-ray albedo) and external
background (due to X-ray diffuse background) was expected to be around 100 keV
\cite{Godet2}. Note that the two missions have similar orbital parameters
(same altitude, but an inclination of $20^\circ$ for {\it ECLAIRs} instead of
$30^\circ$ for SVOM) and the mass model of the {\it ECLAIRs}-CXG camera is
fairly similar to that used on-board SVOM, and the compounds in both mass
models have similar $Z$-values. The mass ratio of the CXG camera between SVOM
and {\it ECLAIRs} is 4.2. We could then expect the internal background of the
SVOM-CXG camera to be roughly four times larger than that computed for the
{\it ECLAIRs}-CXG. This would result in a transition between the internal and
external background to be around 70-90 keV.

\subsection{Preliminary results on the instrument sensitivity}
\label{sensitivity}

\begin{figure}[h]
\begin{center}
\psfig{figure=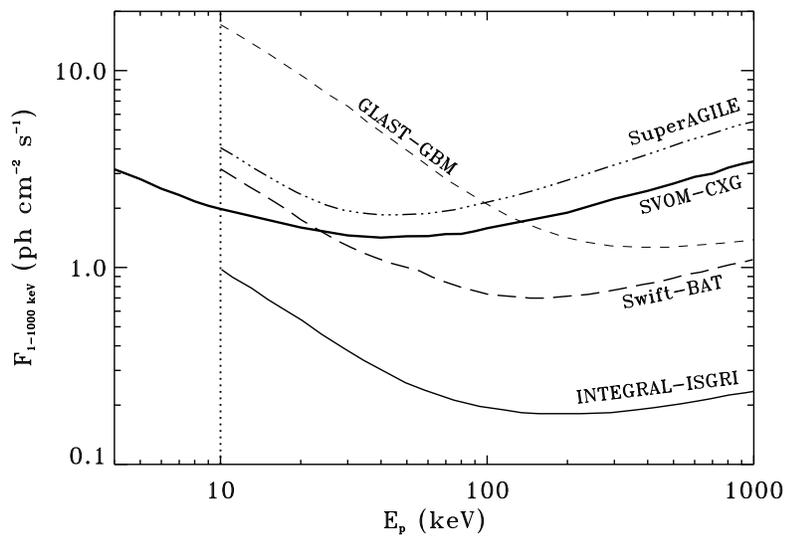,width=11cm}
\caption{Limiting sensitivity $F$ in the $1-10^3$ keV energy range as a
  function of the GRB peak energy, $E_p$. The $F$-values of the CXG were
  computed using the 4-50 keV background level, an integration time of 1.0\,s
  and a $5.5\sigma$ threshold as defined in \cite{Band2}. We considered: i)
  GRB spectra with $\alpha = -0.5$ and $\beta = -2$ ; ii) GRBs located on the
  camera axis.  The sensitivity values for other instruments come from
  \cite{Band3} and were computed from $E_p=10$ keV to $E_p=1000$ keV.}
\label{fig_sens_GRB}
\end{center}
\end{figure}

We used the formalism described in \cite{Band2} to compute the 1-1000 keV
limiting sensitivity of the camera ($F$) for different energy ranges of the
CXG camera as a function of the GRB parameter, $E_p$.  This enables us to
compare directly the expected GRB sensitivity of the CXG with respect to other
present and future GRB instruments \cite{Band3} (see
Fig.~\ref{fig_sens_GRB}). To do so, we considered: i) GRB spectra with $\alpha
= -0.5$ and $\beta = -2$ which are the averaged values derived from the BATSE
$\alpha$ and $\beta$ distributions; ii) GRBs located on the camera axis and an
integration time of 1.0\,s. We used a trigger threshold of $5.5\sigma$.

\begin{figure}
\begin{center}
\psfig{figure=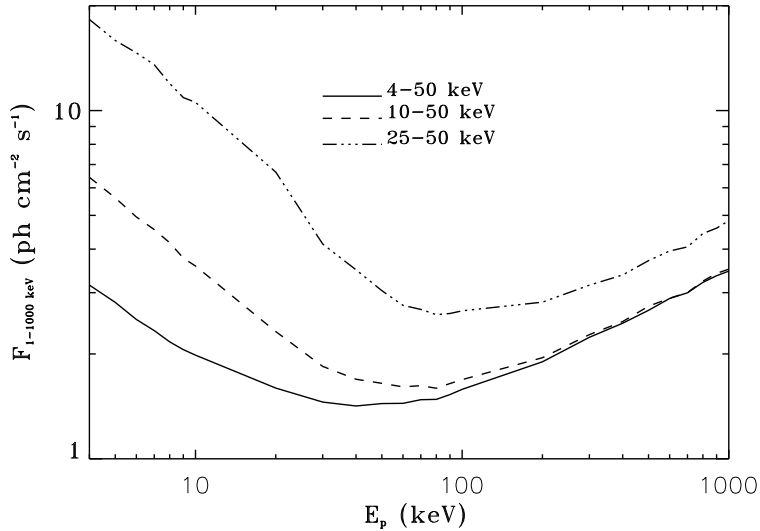,width=11cm} 
\caption{Same as Fig.~\ref{fig_sens_GRB} for three different energy bands:
  4-50 keV (solid line); 10-50 keV (dotted line); 25-50 keV (dashed line). As
  shown, the low energy threshold of the CXG at 4 keV is essential to
  enable the CXG to be more sensitive to high-redshift bursts than the
  {\it Swift}/BAT.}
\label{fig_sens_Thr}
\end{center}
\end{figure}

From Fig.~\ref{fig_sens_GRB}, it appears that the instrument ISGRI on-board
INTEGRAL is the most sensitive instrument. However, its field of view is much
smaller than that of the {\it Swift}-BAT, SuperAGILE, Fermi-GBM and
SVOM-CXG. The CXG will be more sensitive to GRBs with $E_p < 20$ keV than most
of the present and future high-energy instruments except ISGRI, and hence, to
high-redshift GRBs of which the prompt radiation should peak mostly in
X-rays. That is mainly due to the lower energy threshold of the CXG of $\sim
4$ keV (see Fig.~\ref{fig_sens_Thr}) instead of 15 keV for the {\it Swift}-BAT
for instance. However, the effective area of the CXG being less than that of
the {\it Swift}-BAT (5200 cm$^2$), the CXG will be less sensitive than the
{\it Swift}-BAT to GRBs with $E_p > 20$ keV. Taking into account a broader
energy range for the CXG background will only result in a slight improvement
of the CXG sensitivity at higher $E_p$-values. The CXG will also be more
sensitive to GRBs with $E_p < 100$\,keV than the GBM on-board Fermi/GLAST.

\cite{Salvaterra} reviewed the performance of current and future GRB missions
to detect high-redshift GRBs. Their Table~1 shows that the current design of
the SVOM-CXG may be enabled to detect 2-4 GRBs with a redshift larger than 6
per year.

\subsection{Impact of a dead layer in the CdTe detectors on the CXG performance}
\label{dead}

\begin{figure}
\begin{center}
\begin{tabular}{c}
\psfig{figure=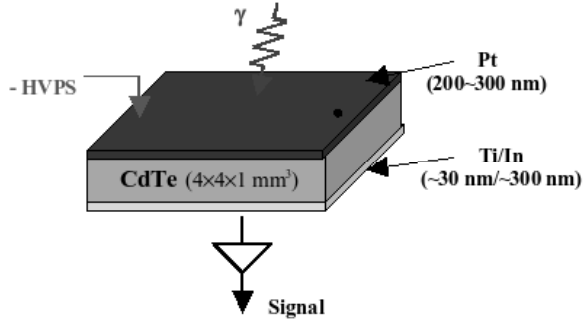,width=8cm} 
\end{tabular}
\caption{Scheme of a Schottky CdTe detector provided by ACRORAD Co. LTD. In
  addition of the CdTe active volume, the detector consists of a cathode
  made of a $200-300$\,nm Platinum (Pt) layer and an anode made of a $\sim
  300$\,nm thick Indium (In) $+$ $\sim 30$\,nm-thick Titanium (Ti) layer. The
  photons illuminating the cathode side of the detectors, the Pt layer acts
  like a dead layer absorbing low-energy photons. }
\label{fig_pixel}
\end{center}
\end{figure}

\begin{figure}[h]
\begin{center}
\begin{tabular}{c}
\psfig{figure=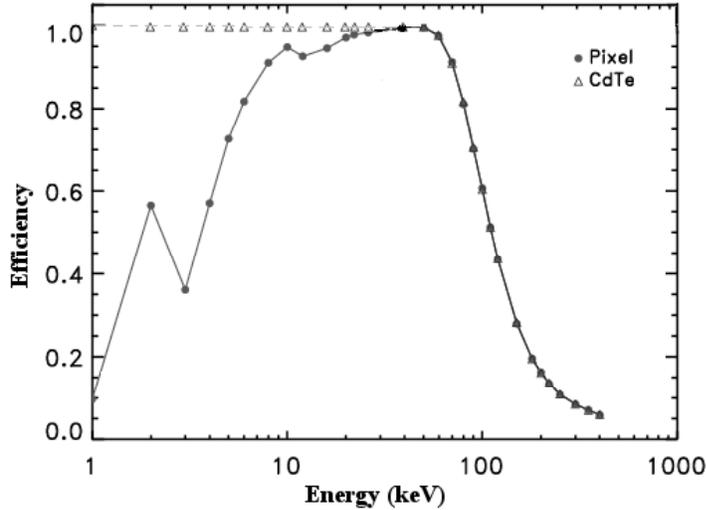,width=9.5cm} 
\end{tabular}
\caption{Comparison of the efficiency of a CdTe pixel without (full triangles)
and with (full circles) a $250$\,nm-thick Pt layer. The Pt layer results in an
absorption of $\sim 40\%$ at 4 keV and $\sim 20\%$ at 6 keV. A full
characterisation of the Pt dead layer thickness for each detector is under
investigation at the CESR lab facility \cite{Remoue2}.}
\label{fig_eff}
\end{center}
\end{figure}

Until then, we assumed all the CdTe detectors in the detection plane were
described as pure CdTe crystals. We then did not model the Schottky contact on
each detector (the anode) made of a $\sim 300$\,nm thick Indium (In) layer $+$
$\sim 30$\,nm-thick Titanium (Ti) layer and the cathode made of a
$200-300$\,nm Platinum (Pt) layer. The Ti layer is added to reinforce the
stability of the In layer.  Figure~\ref{fig_pixel} shows a scheme of a CdTe
detector. The incident photons will penetrate the detectors on the
cathode. The Pt layer acts then like a dead layer on top the active volume of
the detector. Even if this Pt layer is thin, it will absorb a significant
fraction of low-energy photons, since the Pt layer results in an absorption of
$\sim 40\%$ at 4 keV and $\sim 20\%$ at 6 keV (see Fig~\ref{fig_eff}). Above
20 keV, the absorption is less than 5\%. A full characterisation of the Pt
dead layer thickness for each detector is under investigation at the CESR lab
facility \cite{Remoue2}.  In order to quantify the impact of such a dead layer
on the CXG performance, we run Monte-Carlo simulations including a
$250$\,nm-thick Pt dead layer on top each of pixels to model the cathode and a
$300$\,nm thick Indium (In) layer $+$ $30$\,nm-thick Titanium (Ti)
layer to model the anode.

\begin{figure}
\begin{center}
\begin{tabular}{c}
\psfig{figure=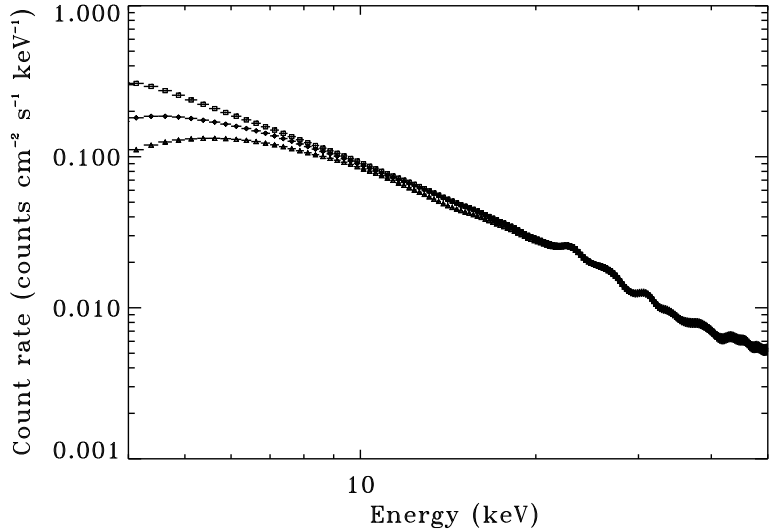,width=11cm} 
\end{tabular}
\caption{Background spectrum in the 4-50 keV energy band illustrating the
  effects of the MLI and Pt layer: (squares) without the MLI and Pt layer;
  (crosses) with the MLI layer and without the Pt layer; (triangles) with
  both.}
\label{fig_bkg_deadlayer}
\end{center}
\end{figure}

Fig.~\ref{fig_bkg_deadlayer} shows the background spectrum of the CXG in the
4-50 keV energy band in three cases: without the MLI layer and the Pt dead
layer (square -- {\scriptsize CASE$_1$}); with the MLI layer, but without the
Pt dead layer (crosses -- {\scriptsize CASE$_2$}); with both (triangles --
{\scriptsize CASE$_3$}). The effects of the MLI and dead layer are clearly
visible at low energy. The introduction of the dead layer in our simulations
results in a 40\% decrease in the background level at 4 keV. The count rate
level is then $1.5$ counts cm$^{-2}$ s$^{-1}$ in the 4-50 keV energy band
(\emph{i.e.} a 12\% reduction with respect to the value found without the Pt
cathode).

\begin{figure}
\begin{center}
\begin{tabular}{c}
\psfig{figure=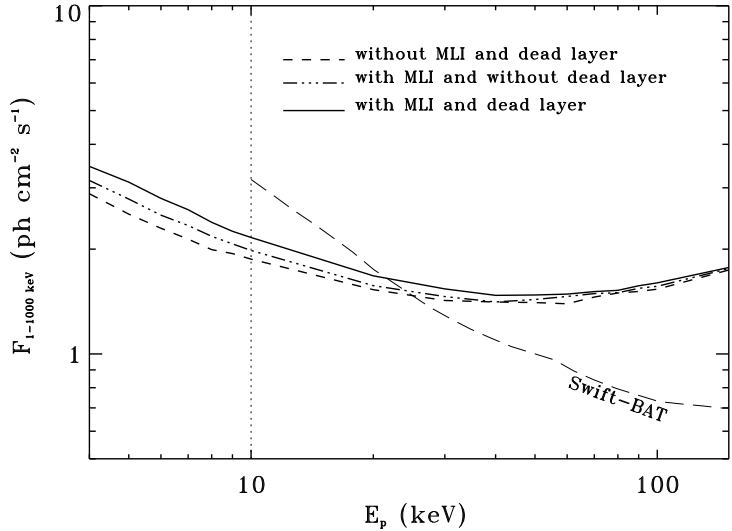,width=11cm} 
\end{tabular}
\caption{Same as Fig.~\ref{fig_sens_GRB}, but assuming 3 different cases:
  (dashed line) without the MLI layer and the Pt dead layer; (dashed and
  dotted line) with the MLI layer and without the Pt dead layer; (solid line)
  with both. Even when taking into account the Pt dead layer, the CXG should
  be more sensitive than the {\it Swift}-BAT to detect GRBs with $E_p < 20$
  keV. }
\label{fig_sens_deadlayer}
\end{center}
\end{figure}

Fig.~\ref{fig_sens_deadlayer} shows the limiting sensitivity in the 1-1000 keV
energy band as a function of the energy peak $E_p$ in different cases.  While
the MLI induces a degradation of the sensitivity by less than 6\% below
$E_p=10$ keV when compared to {\scriptsize CASE$_1$}, the Pt cathode degrades
the limiting sensitivity by less than 10\% below $E_p=10$ keV when compared to
{\scriptsize CASE$_2$}. The impact of the Pt dead layer on the CXG performance
is not significant enough to change the main remarks drawn in
Section~\ref{sensitivity}.

\section{Discussion and Conclusion}

We described in detail our Monte-Carlo simulator to compute the background
spectrum using a given mass model of the coded-mask camera CXG as well as a
model of the spacecraft environment in orbit. We showed that the current
design of the passive shield ensures that in the 4-50 keV imaging energy band,
the background is dominated by the cosmic background. We showed that the MLI
layer and the Pt cathode induce a significant reduction of the background
count rate at low energy.

We demonstrated, using the estimated background level in the 4-50 keV imaging
energy band, that the CXG will be more sensitive to GRBs with $E_p < 20$ keV,
thanks to an expected low-energy threshold around 4 keV, (and therefore
potential high-redshift GRBs) than the {\it Swift}-BAT, presently the
best-designed high-energy GRB imager to date. Computation showed that SVOM
could detect a good sample of high-redshift GRBs during its lifetime.

We also showed that the MLI layer and the Pt cathode induce a non negligible
decrease of the camera sensitivity for values of $E_p < 10$ keV. Investigation
are on-going to see whether or not it is possible further to reduce the
thickness of the MLI layer.

Further improvements will be also made in our simulator to refine results
concerning: i) the internal background due the material activation in orbit;
ii) the evolution of the background level near and in the SAA. We will also
investigate the impact on the CXG performance of a higher contribution of the
cosmic background below 10 keV.

We are also investigating the benefit to use a coded mask with an aperture of
40\%, instead of 30\% as presented in this paper, in order to increase the
fraction of short GRBs detectable by the CXG without degrading too much the
sensitivity of the camera.

The imaging performance of the CXG will be reviewed in a forthcoming paper.

\medskip

{\bf Acknowledgements}

OG gratefully acknowledges STFC funding and the CESR that hosted him during the
writing of this paper, and where most of this work was done during his PhD.

\end{document}